# Markov State Models of Gene Regulatory Networks

Brian K. Chu[1], Margaret J. Tse[1], Royce R. Sato[1], And Elizabeth L. Read[1,2]

[1]*Department of Chemical Engineering and Materials Science and* [2]*Department of Molecular Biology and Biochemistry, University of California Irvine, Irvine, California*

Email addresses: chubk@uci.edu, tsem1@uci.edu, rrysato@gmail.com, elread@uci.edu

**ABSTRACT:**

**BACKGROUND:** Gene regulatory networks with dynamics characterized by multiple stable states underlie cell fate-decisions. Quantitative models that can link molecular-level knowledge of gene regulation to a global understanding of network dynamics have the potential to guide cell-reprogramming strategies. Networks are often modeled by the stochastic Chemical Master Equation, but methods for systematic identification of key properties of the global dynamics are currently lacking.

**METHODS:** We present a method for analyzing global dynamics of gene networks using the Markov State Model framework, which utilizes a separation-of-timescales based clustering method to obtain a coarse-grained state transition graph that approximates global gene network dynamics.

**RESULTS:** The method identifies the number, phenotypes, and lifetimes of long-lived states for a set of common gene regulatory network models. Application of transition path theory to the constructed Markov State Model decomposes global dynamics into a set of dominant transition paths and associated relative probabilities for stochastic state-switching.

**CONCLUSIONS:** In this proof-of-concept study, we found that the Markov State Model provides a general framework for analyzing and visualizing stochastic multistability and state-transitions in gene networks. Our results suggest that this framework—adopted from the field of atomistic Molecular Dynamics—can be a useful tool for quantitative Systems Biology at the network scale.

**BACKGROUND**

Gene regulatory networks (GRNs) often have dynamics characterized by multiple attractor states. This multistability is thought to underlie cell fate-decisions. According to this view, each attractor state accessible to a gene network corresponds to a particular pattern of gene expression, i.e., a cell phenotype. Bistable network motifs with two possible outcomes have been linked to binary cell fate-decisions, including the lysis/lysogeny decision of bacteriophage lambda[1], the maturation of frog oocytes[2] and a cascade of branch-point decisions in mammalian cell development (reviewed in [3]). Multistable networks with three or more attractors have been proposed to govern diverse cell fate-decisions in tumorigenesis [4], stem cell differentiation and reprogramming [5]–[7], and helper T cell differentiation[8]. More generally, the concept of a rugged, high-dimensional epigenetic landscape connecting every possible cell type has emerged [9]–[11]. Quantitative models that can link molecular-level knowledge of gene regulation to a global understanding of network behavior have the potential to guide rational cell-reprogramming strategies. As such, there has been growing interest in the development of theory and computational methods to analyze global dynamics of multistable gene regulatory networks.

Gene expression is inherently stochastic[1], [12]–[14], and fluctuations in expression levels can measurably impact cell phenotypes and behavior. Numerous examples of stochastic phenotype transitions have been discovered, which diversify otherwise identical cell-populations. This spontaneous state-switching has been found to promote survival of microorganisms or cancer cells in fluctuating environments[15]–[17], prime cells to follow alternate developmental fates in higher eukaryotes [18], [19], and generate sustained heterogeneity (mosaicism) in a homeostatic mammalian cell population[20]. These findings have motivated theoretical studies of stochastic state-switching in gene networks, which have shed light on network parameters and topologies that promote the stability (or instability) of a given network state[20]. Characterizing the global stability of states accessible to a network is akin to quantification of the "potential energy" landscape of a network. Particularly, with the advent of stem-cell reprogramming techniques, there has been renewed interest in a quantitative reinterpretation of Waddington's classic epigenetic landscape[21], in terms of underlying regulatory mechanisms[10], [22].

A number of mathematical frameworks exist for modeling and analysis of stochastic gene regulatory network (GRN) dynamics (reviewed in [23], [24]), including probabilistic Boolean Networks, Stochastic Differential Equations, and stochastic biochemical reaction networks (i.e., Chemical Master Equations). Of these, the Chemical Master Equation (CME) approach is the most complete, in that it treats all biomolecules in the system as discrete entities, fully accounts for stochasticity due to molecular-level fluctuations, and propagates dynamics according to chemical rate laws. The CME is analytically intractable, but trajectories can be simulated by Monte Carlo methods such as the Stochastic Simulation Algorithm (SSA)[25]. Alternatively, methods for reducing the dimensionality of the CME, enabling numerical approximation of network behavior by matrix methods, have been developed[26]–[30].

Analysis of multistability and global dynamics of discrete, stochastic GRN models remains challenging. Multistability is generally assessed by plotting multi-peaked steady-state probability distributions (obtained either from long stochastic simulations [5], [31], [32] or from approximate CME solutions[30], [33], [34]), projected onto one or two user-specified system coordinates. However, even small networks generally have more than two dimensions along which dynamics may be projected, meaning that inspection of steady-state distributions for a given projection may underestimate multistability in a network. For example, the state-space of a GRN may comprise different activity-states of promoters and regulatory sites on DNA, the copy-number of mRNA transcripts and encoded proteins, and the activity- or multimer-states of multiple regulatory molecules or proteins. Furthermore, while steady-state distributions give a global view of system behavior, they do not directly yield dynamic information of interest, such as the lifetimes of attractor states.

In this paper, we present an approach for analyzing multistable dynamics in stochastic GRNs based on a spectral clustering method widely applied in Molecular Dynamics[35]–[37]. The output of the approach is a Markov State Model (MSM)—a coarse-grained model of system dynamics, in which a large number of system states (i.e., "microstates") is clustered into a small number of metastable (that is, relatively long-lived) "macrostates", together with the conditional probabilities for transitioning from one macrostate to another on a given timescale. The MSM approach identifies clusters based on separation of timescales, i.e., systems with multistability exhibit relatively fast transitions among microstates within attractor basins and relatively slow inter-basin transitions. By neglecting fast transitions, the size of the system is vastly reduced. Based on its utility for visualization and analysis of Molecular Dynamics, the potential application of the MSM framework to diverse dynamical systems, including biochemical networks, has been discussed[38].

Biochemical reaction networks present an unexplored opportunity for the MSM approach. Herein, we applied the method to small GRN motifs and analyzed their global dynamics using two frameworks: the quasipotential landscape (based on the log-transformed stationary probability distribution), and the MSM. The MSM approach distilled network dynamics down to the essential stationary and dynamic properties, including the number and identities of stable phenotypes encoded by the network, the global probability of the network to adopt a given phenotype, and the likelihoods of all possible stochastic phenotype transitions. The method revealed the existence of network states and processes not readily apparent from inspection of quasipotential landscapes. Our results demonstrate how MSMs can yield insight into regulation of cell phenotype stability and reprogramming. Furthermore, our results suggest that, by delivering systematic coarse-graining of high-dimensional (i.e., many-species) dynamics, MSMs could find more general applications in Systems Biology, such as in signal-transduction, evolution, and population dynamics.

## METHODS

*Gene Regulatory Network Motifs*

We studied two common GRN motifs that are thought to control cell fate-decisions. The full lists of reactions and associated rate parameters for each network are given in the Supplement. Both motifs consist of two mutually-inhibiting genes, denoted by *A* and *B*. In the Exclusive Toggle Switch (ETS) motif, each gene encodes a transcription factor protein; the protein forms homodimers, which are capable of binding to the promoter of the competing gene, thereby repressing its expression. One DNA-promoter region controls the expression of both genes; when a repressor is bound, it excludes the possibility of binding by the repressor encoded by the competing gene. Therefore, the promoter can exist in three possible binding configurations, $P_{00}$, $P_{10}$, and $P_{01}$, denoting the unbound, $a_2$-bound, or $b_2$-bound states, respectively. Production of new protein molecules (including all processes involved in transcription, translation, and protein synthesis) occurs at a constant rate, which depends on the state of the promoter. When the gene is repressed, the encoded protein is produced at a low rate, denoted $g_0$. When the gene is not repressed, protein is produced at a high rate, $g_1$. For example, when the promoter state is $P_{10}$ the *a* protein is produced at rate $g_1$, and the *b* protein is produced at $g_0$. When the promoter is unbound, neither gene is repressed, causing both proteins to be produced at rate $g_1$.

In the Mutual Inhibition/Self-Activation (MISA) motif, each homodimeric transcription factor also activates its own expression, in addition to repressing the other gene. The *A* and *B* genes are controlled by separate promoters, and each promoter can be bound by repressor and activator simultaneously. Therefore, the *A*-promoter can exist in four possible states, $A_{00}$, $A_{10}$, $A_{01}$ and $A_{11}$, denoting unbound, $a_2$-activator bound, $b_2$-repressor bound, and both transcription factors bound, respectively (and similarly for the *B*-promoter). Proteins are produced at rate $g_1$ only when the activator is bound and the repressor is unbound. For example, the $A_{10}$ promoter state allows *a* protein to be produced at $g_1$. The other three *A* promoter states result in *a* protein being produced at rate $g_0$. Similarly, the rate of *b* protein production depends only on the binding configuration of the *B*-promoter. In both the ETS and MISA networks, protein dimerization is assumed to occur simultaneously with binding to DNA. All rate parameters are given in Tables S1-2.

*Chemical Master Equation*

The stochastic dynamics are modeled by the discrete, Markovian Chemical Master Equation, which gives the time-evolution of the probability to observe the system in a given state over time. In vector-matrix form, the CME can be written

$$\frac{d\mathbf{p}(\mathbf{x},t)}{dt} = \mathbf{K}\mathbf{p}(\mathbf{x},t)$$

where $p(x, t)$ is the probability over the system state-space at time $t$, and $K$ is the reaction rate-matrix. The off-diagonal elements $K_{ij}$ give the time-independent rate of transitioning from state $x_i$ to $x_j$, and the diagonal elements are given by $K_{ii} = -\sum_{j \neq i} K_{ji}$. We assume a well-mixed system of reacting species, and the state of the system is fully specified by $x \in \mathbb{N}^S$, a state-vector containing the positive-integer values of all $S$ molecular species/configurations. We hereon denote these state-vectors as "microstates" of the system. In the ETS network, $x = [n_A, n_B, P_{ab}]$, where $n_A$ is the copy-number of $a$ molecules (protein monomers expressed by gene $A$, and similar for $B$), and $P_{ab}$ indexes the promoter binding-configuration. In the MISA network, $x = [n_A, n_B, A_{ab}, B_{ba}]$, which lists the protein copy numbers and promoter configuration-states associated with both genes.

The reaction rate matrix $K \in \mathbb{R}^{N \times N}$ is built from the stochastic reaction propensities (Supplement Eq. 1)[39], for some choice of enumeration over the state-space with $N$ reachable microstates. In general, if a system of $S$ molecular species has a maximum copy number per species of $n_{max}$, then $N \sim n_{max}^S$. To enumerate the system state-space, we neglect microstates with protein copy-numbers larger than a threshold value, which exceeds the maximum steady-state gene expression rate, $g_1/k$ (where $g_1$ is the maximum production rate of protein and $k$ is the degradation rate), as these states are rarely reached. This truncation of the state-space introduces a small approximation error, which we calculate using the Finite State Projection method [40] (Fig. S1).

*Stochastic Simulations*

Stochastic simulations were performed according to the SSA method [25], implemented by the software package StochKit2 [41].

*Quasipotential Landscape*

The steady-state probability $\boldsymbol{\pi}(\mathbf{x})$ over *N* microstates is obtained from $\mathbf{K}$ as the normalized eigenvector corresponding to the zero-eigenvalue, satisfying $\mathbf{K}\boldsymbol{\pi}(\mathbf{x})=\mathbf{0}$[42]. Quasipotential landscapes were obtained from $\boldsymbol{\pi}(\mathbf{x})$ using a Boltzmann definition, $U(\mathbf{x}) = -\ln(\boldsymbol{\pi}(\mathbf{x}))$[22]. All matrix calculations were performed with MATLAB[43].

*Markov State Models: Mathematical Background*

The last 15 years have seen continual progress in development of theory, algorithms, and software implementing the MSM framework. We briefly summarize the theoretical background here; the reader is referred to other works (e.g., [44]–[48]) for more details.

The MSM is a highly coarse-grained projection of system dynamics over *N* microstates onto a reduced space of selected size $C$ (generally, $C \ll N$). The $C$ states in the projected dynamics are constructed by clustering together microstates that experience relatively fast transitions among them. The $C$ clusters, also called "almost invariant aggregates"[49], are hereon denoted "macrostates".

The MSM approach makes use of Robust Perron Cluster Analysis [50] (PCCA+), a spectral clustering algorithm that takes as input a row-stochastic transition matrix, $\mathbf{T}(\tau)$ which gives the conditional probability for the system to transition between each pair of microstates within a given lagtime $\tau$. The lagtime determines the time-resolution of the model, as expressed by the transition matrix. Off-diagonal elements $T_{ij}$ give the probability of the system to transition to microstate $j$ within $\tau$, given that it was initialized in $i$. Diagonal elements $T_{ii}$ give the conditional probability to remain in microstate $i$ over the $\tau$ interval, and thus rows sum to 1. $\mathbf{T}(\tau)$ is directly obtained from the reaction rate matrix by [51]:

$$\mathbf{T}(\tau) = \exp(\tau \mathbf{K}^{\mathrm{T}}),$$

(where $\exp$ denotes the matrix exponential). The evolution of the probability over discrete intervals of $\tau$ is given by the Chapman-Kolmogorov equation,

$$\boldsymbol{p}^T(\mathbf{x}, t + k\tau) = \boldsymbol{p}^T(\mathbf{x}, t)\mathbf{T}^k(\tau).$$

For an ergodic system (i.e., any state in the system can be reached from any other state in finite time), $\mathbf{T}(\tau)$ will have one largest eigenvalue, the Perron root, $\lambda_1 = 1$. The stationary probability is then given by the normalized left-eigenvector corresponding to the Perron eigenvalue,

$$\boldsymbol{\pi}^{\mathrm{T}}(\mathbf{x})\mathbf{T}(\tau) = \boldsymbol{\pi}^{\mathrm{T}}(\mathbf{x}).$$

If the system exhibits multistability, then the dynamics can be approximately separated into fast and slow processes, with fast transitions occurring between microstates belonging to the same metastable macrostate, and slow transitions carrying the system from one macrostate to another. Then $\mathbf{T}(\tau)$ is nearly decomposable, and will exhibit an almost block-diagonal structure (for an appropriate ordering of microstates) with $C$ nearly uncoupled blocks. In this case, the eigenvalue spectrum of $\mathbf{T}(\tau)$ shows a cluster of $C$ eigenvalues near $\lambda_1 = 1$, denoting $C$ slow processes (including the stationary process), and for $i > C$, $\lambda_i \ll \lambda_C$, corresponding to rapidly decaying processes. The system timescales can be computed from the eigenvalue spectrum according to $t_i = -\tau/\ln|\lambda_i(\tau)|$.

The PCCA+ algorithm obtains fuzzy membership vectors $\boldsymbol{\chi} = [\chi_1, \chi_2, \ldots, \chi_C] \in \mathbb{R}^{N \times C}$, which assigns microstates $i \in \{1, \ldots, N\}$ to macrostates $j \in \{1, \ldots, C\}$ according to grades (i.e., probabilities) of membership, $\chi_j(i) \in [0,1]$. The membership vectors satisfy the linear transformation:

$$\boldsymbol{\chi} = \boldsymbol{\psi} \mathbf{B}$$

Where $\boldsymbol{\psi} = [\psi_1, \ldots, \psi_C]$ is the $N \times C$ matrix constructed from the $C$ dominant right-eigenvectors of $\mathbf{T}(\tau)$, and $\mathbf{B}$ is a non-singular matrix whose elements are determined by an optimization procedure. The original PCCA method [49] used the sign structure of the eigenvectors to identify almost invariant aggregates, in an optimization procedure with the objective of maximizing metastability via the trace of the coarse-grained matrix. A new optimization algorithm was introduced in a method known as PCCA+ [50], which improved numerical robustness. The results of this paper were generated using the PCCA+ implementation of MSMBuilder2 [52].

*Construction of Markov State Models and Pathway Decomposition*

The PCCA+ algorithm generates a fuzzy discretization. We convert fuzzy values into a so-called "crisp" partitioning of $N$ states into $C$ clusters, which entirely partitions the space with no overlap, by assigning $\chi_j^{crisp}(i) \in \{0,1\}$. That is, $\chi_j^{crisp}(i) = 1$ if the $j$th element of the row vector $\chi(i)$ is maximal, and 0 otherwise. Transition probabilities are estimated over the $C$ coarse-grained sets by summing over the fluxes, or equivalently[53]:

$$\widetilde{\mathbf{T}}(\tau) = \widetilde{\boldsymbol{D}}^{-1} \boldsymbol{\chi}^T \boldsymbol{D} \mathbf{T}(\tau) \boldsymbol{\chi},$$

where $\widetilde{\mathbf{T}}(\tau) \in \mathbb{R}^{C \times C}$ is the coarse-grained Markov State Model and $\mathbf{D}$ is the diagonal matrix obtained from the stationary probability vector, $\mathbf{D} = \text{diag}(\pi_1, \dots, \pi_N)$. The coarse-grained probability $\widetilde{\boldsymbol{\pi}}(\mathbf{x})$ is obtained by $\widetilde{\boldsymbol{\pi}}(\mathbf{x}) = \boldsymbol{\chi}^{\text{T}} \boldsymbol{\pi}(\mathbf{x})$, and $\widetilde{\mathbf{D}} = \text{diag}(\widetilde{\pi}_1, \dots, \widetilde{\pi}_C)$.

The Markov State Model is visualized using the PyEmma 2 plotting module[44], where the magnitude of the transition probabilities and steady state probabilities are represented by the thickness of the arrows and size of the circles, respectively.

Upon construction of the Markov State Model, transition-path theory[54]–[56] was applied in order to compute an ensemble of transition paths connecting two states of interest, along with their relative probabilities. This was achieved by applying a pathway decomposition algorithm adapted from Noe, et al. in a study of protein folding pathways[56] (details in Supplement).

**RESULTS**

*Eigenvalues and Eigenvectors of the stochastic transition matrix reveal slow dynamics in gene networks*

In order to explore the utility of the MSM approach for analyzing global dynamics of gene networks, we studied common motifs that control lineage decisions. The MISA network motif (Fig. 1A, Supplement, and Methods) has been the subject of previous theoretical studies and is thought to appear in a wide variety of binary fate-decisions. [57]–[59]. In the network model, the *A/B* gene pair represents known antagonistic pairs such as Oct4/Cdx2, PU.1/Gata1, and GATA3/T-bet, which control lineage decisions in embryonic stem cells, common myeloid progenitors, and naïve T-helper cells, respectively [9], [60], [61]. In general, a particular cell lineage will be associated with a phenotype in which one of the genes is expressed at a high level, and the other is expressed at a low (repressed) level. The MISA network has been reported to have up to four attractors[4], [5], corresponding to the *A/B* gene pair expression combinations Lo/Lo, Lo/Hi, Hi/Lo, and Hi/Hi. We computed the probability and quasipotential landscape of the MISA network. For a symmetric system with sufficiently balanced rates of activator and repressor binding and unbinding from DNA, four peaks (attractor basins) can be distinguished in the steady state probability (quasipotential) landscape, plotted as a function of protein *a* copy number vs. protein *b* copy number (Fig. 1A,B).

The Markov State Model framework has been applied in studies of protein folding, where dynamics occurs over rugged energetic landscapes characterized by multiple long-lived states (reviewed in [46], [62]). Therefore, we reasoned that the approach could be useful for studying global dynamics of multistable GRNs. The method identifies the slowest system processes based on the dominant eigenvalues and eigenvectors of the stochastic transition matrix, $\mathbf{T}(\tau)$, which gives the probability of the system to transition from every possible initial state to every possible destination state within lagtime $\tau$ (with $\tau$ having units of $k^{-1}$ and $k$ being the rate of protein degradation). Inspection of the eigenvalue spectrum of $\mathbf{T}(\tau = 5)$ for the MISA network in Fig. 1B reveals four eigenvalues near 1 followed by a gap, indicating four system processes that are slow on this timescale. Decreasing $\tau$ to 0.5 reveals a step-structure in the eigenvalue spectrum, suggesting a hierarchy of system timescales. The timescales are related to the eigenvalues according to $t_i = -\tau/\ln|\lambda_i(\tau)|$. The Perron eigenvalue $\lambda_1 = 1$ is associated with the stationary (infinite time) process, and the lifetimes $t_2$ through $t_5$ are computed to be {95.6, 49.4, 30.8, 2.6} (in units of $k^{-1}$). Thus, the first gap in the eigenvalue spectrum arises from a more than ten-fold separation in timescales between $t_4$ and $t_5$. The original PCCA method[49] used the sign structure of the eigenvectors to assign cluster memberships. Plotting the left-eigenvectors corresponding to the four dominant eigenvalues in the MISA network is instructive: the stationary landscape is obtained from the first eigenvector ($\phi_1$), which is positive over all microstates, while the opposite-sign regions in $\phi_2$, $\phi_3$, $\phi_4$ reveal the nature of the slow processes (Fig. 1D). An eigenvector with regions of opposite sign corresponds to an exchange between those two regions (in both directions, since eigenvectors are sign-interchangeable). For example, the slowest process corresponds to exchange between the $a > b$ and $b > a$ regions of state-space, i.e., switching between *B*-gene dominant and *A*-gene-dominant expression states. Eigenvectors $\phi_3$ and $\phi_4$ show that somewhat faster timescales are associated with exchange in and out of the Lo/Lo and Hi/Hi basins.

*The Markov State Model approach identifies multistability in GRNs*

**Reduced Models of the MISA Network.** The MSM framework utilizes a clustering algorithm known as PCCA+ (see Methods and Supplement) to assign every microstate in the system to a macrostate (i.e., a cluster of microstates) based on the slow system processes identified by the eigenvectors and eigenvalues of $\mathbf{T}(\tau)$. Applying the PCCA+ algorithm to the MISA network for the parameter set of Fig. 1 resulted in a mapping from $N = 15{,}376$ ($31 \times 31 \times 4 \times 4$) microstates onto $C = 4$ macrostates (Fig. 2). The $N$ microstates were first enumerated by accounting for all possible system configurations with $0 \leq a \leq 30$ and $0 \leq b \leq 30$. This enumeration assumes a negligible probability for the system to ever exceed 30 copies of either protein, which introduces a small approximation error of $1E-5$ (details in Fig. S1). Because the promoters of each gene can take four possible configurations—that is, two binding sites (for the repressor and activator) that can be either bound or unbound—a total of 16 gene configuration states are possible, giving $N = 15,376$ enumerated microstates. Quasipotentials calculated from a long brute force simulation and from $\phi_1$ showed agreement (Fig. S2). For this parameter set, the four macrostates obtained correspond to the visible peaks (basins) in the probability (quasipotential) landscape (Fig. 2A). The average expression levels of proteins in each macrostate indicate the four distinct cell phenotypes (Lo/Lo, Lo/Hi, Hi/Lo, Hi/Hi). The representative gene promoter configurations for each macrostate are shown (Fig. 2B). However, for each macrostate cluster there are other possible gene promoter configurations present with lower steady-state probability that are not shown, since every enumerated microstate is assigned to a macrostate.

**Parameter-dependence of landscapes and MSMs.** To determine whether the MSM approach can robustly identify gene network macrostates, we applied it over a range of network parameters by varying the repressor unbinding rate $f_r$ (all parameters defined in Table S1). Increasing $f_r$ relative to other network parameters modulates the quasipotential landscape by increasing the probability of the Hi/Hi phenotype, in which both genes express at a high level simultaneously (Fig 3B). This occurs as a result of weakened repressive interactions, since the lifetimes of repressor occupancy on promoters are shortened when $f_r$ is increased. The eigenvalue spectra show a corresponding shift: when $f_r = 1E-3$, four dominant eigenvalues are present. When $f_r$ is increased to $f_r = 1$, the largest visible gap in the eigenvalue spectrum shifts to occur after the first eigenvalue ($\lambda =$ 1), indicating loss of multistability on the timescale of $\tau$ (here, $\tau = 5$) (Fig. 3A). Correspondingly, for this parameter set, the landscape shows only a single visible Hi/Hi basin.

The PCCA+ algorithm seeks $C$ long-lived macrostates, where $C$ is user-specified. We constructed Markov State Models for the MISA network over varying $f_r$, specifying four macrostates. The MSMs are shown graphically in Fig 3D. The sizes of the circles are proportional to the relative steady-state probability of the macrostate, and the thickness of the directed edges are proportional to the relative transition probability within $\tau$. In agreement with the landscapes, the MSMs over this parameter regime show increasing probability of the Hi/Hi state, as a result of an increasing ratio of transition probability "into" versus "out of" the Hi/Hi state. The locations of the clusters in the state-space (according to 50% probability contours) do not change appreciably. The choice of lagtime $\tau$ sets the timescale on which metastability is defined in the system. However, in practice, the PCCA+ seeks an assignment of $C$ clusters regardless of whether $C$ metastable states exist in the system on the $\tau$ timescale, and the resulting aggregated macrostates are generally invariant to $\tau$. Thus, for $f_r = 1$, the algorithm locates four macrostates, although the (low-probability) Hi/Lo, Lo/Lo, and Lo/Hi macrostates are likely to experience transitions away, into the Hi/Hi macrostate, within $\tau$. These low-probability states appear in the landscape as shoulders on the outskirts of the Hi/Hi basin. Overall, Fig. 3 demonstrates that, for this parameter regime, the quasipotential landscape and the MSM yield similar information on the global system dynamics in terms of the number and locations of attractor states, and their relative probabilities as a function of the unbinding rate parameter $f_r$. The MSM further provides quantitative information on the probabilities (and thus timescales) of transitioning between each pair of macrostates.

**MSM identifies purely stochastic multistability.** Multistability in gene networks is often analyzed within an ordinary differential equation (ODE) framework, by graphical analysis of isoclines and phase portraits, or by linear stability analysis [4], [8]. ODE models of gene networks treat molecular copy numbers (i.e,. proteins, mRNAs) as continuous variables and apply a quasi-steady-state approximation to neglect explicit binding/unbinding of proteins to DNA. Previous studies have shown that such ODE models can give rise to landscape structures that are qualitatively different from those of their corresponding discrete, stochastic networks. For example, multistability in an ODE model of the genetic toggle switch requires cooperativity—i.e., multimers of proteins must act as regulators of gene expression[63]. However, it was found that monomer repressors are sufficient to give bistability in a stochastic biochemical model[64], [65]. We compared the dynamics of the monomer ETS network (shown schematically in Fig. 4A) as determined by analysis of the ODEs, along with the corresponding stochastic quasipotential landscape and the MSM. In a small-number regime, the ODEs predict monostability (Fig. 4C), while the stochastic landscape shows tristability—that is, three basins corresponding to the Hi/Lo, Hi/Hi, and Lo/Hi expressing phenotypes (Fig. 4A). This discrepancy has been shown to occur in systems with small number effects, i.e., extinction at the boundaries[65].

The MSM approach identifies three metastable macrostates for the monomer ETS in this parameter regime, as seen in the eigenvalue spectrum, which shows a gap after the third index. The reduced Markov State Model constructed for this network thus reduces the system from $N = 7,803$ ($51 \times 51 \times 3$) microstates to $C = 3$ macrostates (Fig. 4B), corresponding to the same Hi/Lo, Hi/Hi, and Lo/Hi attractor phenotypes seen in the quasipotential landscape. Figure 4 demonstrates that the MSM approach can accurately identify purely stochastic multistability in systems where continuous models predict only a single stable fixed-point steady state.

*Analyzing global gene network dynamics with the Markov State Model*

**MSM provides good approximation to relaxation dynamics from a given initial configuration.** Figs. 1-4 demonstrate the utility of the MSM approach for analyzing stationary properties of networks—that is, for identifying the number and locations of multiple attractors at steady state. Additionally, the MSM can be used to make dynamic predictions about transitions among macrostates. Dynamics for either the "full" transition matrix (with all system states enumerated up to a maximum protein copy number) or reduced transition matrix (i.e., the MSM) is propagated according to the Chapman-Kolmogorov equation (see Methods and Supplement). We sought to determine the accuracy of the dynamic predictions obtained from the MSM. Applying the methods proposed by Prinz, et al.([45]) (details in Supplement), we compared the dynamics propagated by the fully enumerated transition matrix $T(\tau)$, which is then projected onto the coarse-grained macrostates, to the dynamics of the coarse-grained system propagated by $\widetilde{T}(\tau)$ (i.e., the MSM). We thus computed the error in dynamics of relaxation out of a given initial system configuration. The system relaxation from a given initial microstate can also be computed by running a large number of brute force SSA simulations. Relaxation dynamics for the full, brute-force, and reduced MSM methods, applied to the MISA with $f_r = 1E - 2$, all show good agreement (Fig. 5 A,B, and C). The error computed between the reduced MSM vs. full dynamics (i.e., $\widetilde{T}(\tau)$ vs $T(\tau)$), is maximally $7.8E - 3$, varies over short times, and decreases continuously after time $t = 140$. Alternatively, the error of the MSM can be quantified by comparing the autocorrelation functions of the MSM and brute force simulation[51], [66]. In Figure S3, we show that the derived autocorrelation functions of the MSM and brute force, and the relaxation constants $\tau_r$, which describes the amount of time to reach equilibrium, are close in value ($\tau_r = 1E3$, for the MSM, and $\tau_r = 1.1E3$ for the brute force). Overall, these results demonstrate that the most accurate predictions of the coarse-grained MSM can be obtained on long timescales, but dynamic approximations with reasonable accuracy can also be obtained for short timescales.

**Parameter-dependence of MSM error.** The accuracy of the MSM dynamic predictions depends on whether inter-macrostate transitions can be treated as memory-less hops. Previous theoretical studies of gene network dynamics found that the height of the barrier separating phenotypic states, and the state-switching time associated with overcoming the barrier, depends on the rate parameters governing DNA-binding by the protein regulators[5], [6], [65], [67]. We reasoned that a larger timescale separation between intra- and inter-basin transitions (corresponding to a larger barrier height separating basins) should result in higher accuracy of the MSM approximation. Thus, we hypothesized that the accuracy of the MSM dynamic predictions should depend on the DNA-binding and unbinding rate parameters. We demonstrated this using the dimeric ETS motif, by computing the error of the MSM approximation for a range of repressor unbinding rates $f$. We varied the binding kinetics without changing the overall relative strength of repression, by varying $f$ together with the repressor binding rate $h$, to maintain a constant binding equilibrium ($X_{eq} = \frac{f}{h} = 100$). By varying $f$ and $h$ in this way over eight orders of magnitude, we found that the barrier height and timescale of the slowest system process ($t_2$) had a non-monotonic dependence on the binding/unbinding parameters. Thus, the fastest inter-phenotype switching was observed in the regime with intermediate binding kinetics, in agreement with previous work[58]. The system also exhibits a shift from three visible basins in the quasipotential landscape in the small $f$ regime to two basins in the large $f$ regime. We performed clustering by selecting $C = 2$ (dashed lines, Fig. 6) and $C = 3$ clusters (solid lines, Fig. 6), and computed the total error over all choices of system initialization, as well as the error associated with relaxation from a particular system microstate. In general, we find that the 3-state MSM approximation is more accurate than the 2-state partitioning. The 3-state MSM dynamic predictions are highly accurate when the DNA-binding/unbinding kinetics is slow. As such, in this regime the Markovian assumption of memory-less transitions between the three phenotypic states is most accurate. As hypothesized, the accuracy of the MSM approximation is lowest (highest error) when the lifetime $t_2$ is shortest (intermediate regime, $f = 1$), and the error decreases modestly with further increase in $f$ (i.e., increase in $t_2$).

**Decomposition of state-transition pathways in gene networks using the MSM framework.** Quantitative models of gene network dynamics can shed light on transition paths connecting phenotypic states. The MSM approach coupled with transition path theory[68]–[70] enables decomposition of all major pathways linking initial and final macrostates of interest. This type of pathway decomposition has previously shed light on mechanisms of protein folding[56]. We demonstrate this pathway decomposition on the MISA network, by computing the transition paths linking the polarized *A*-dominant (Hi/Lo) and *B*-dominant (Lo/Hi) phenotypes. Multiple alternative pathways linking these phenotypes are possible: for the 4-state coarse-graining, the system can alternatively transit through the Hi/Hi or Lo/Lo phenotypes when undergoing a stochastic state-transition from one polarized phenotype to the other. Not all possible paths are enumerated since only transitions with net positive fluxes are considered (see Equation S18). The hierarchy of pathway probabilities for successful transitions depends on the kinetic rate parameters (Fig. 7A). It could be tempting to intuit pathway intermediates based on visible basins in the quasipotential landscape. However, we found that the steady-state probability of an intermediate macrostate (i.e., the Hi/Hi or Lo/Lo states) does not accurately predict if it serves as a pathway intermediate for successful transitions, because parameter regimes are possible in which successful transitions are likely to transition through intermediates with high potential/low probability (Fig. 7C). This occurs because the relative probability of transiting through one intermediate macrostate versus another is based on the balance of probabilities for entering and exiting the intermediate: intermediate states that can be easily reached—but not easily exited—as a result of stochastic fluctuations can act as "trap" states. Therefore, it is shown that the pathway probability cannot be inferred from the steady state probability of the intermediates alone.

**MSMs can be constructed with different resolutions of coarse-graining.** The eigenvalue spectrum of the MISA network shows a step-structure, with nearly constant eigenvalue clusters separated by gaps. These multiple spectral gaps suggest a hierarchy of dynamical processes on separate timescales. A convenient feature of the MSM framework is that it can build coarse-grained models with different levels of resolution by PCCA+, in order to explore such hierarchical processes. We applied the MSM framework to a MISA network with very slow rates of DNA-binding and unbinding ($f_r = 1E-4, h_r = 1E-6$), comparing the macrostates obtained from selecting $C = 4$ versus $C = 16$ clusters. For $\mathbf{T}(\tau = 1)$, a prominent gap occurs in the eigenvalue spectrum between $\lambda_{16}$ and $\lambda_{17}$, corresponding to an almost 30-fold separation of timescales between $t_{16} = 27.8$ and $t_{17} = 0.99$(Fig. 8A). Applying PCCA+ with $C = 16$ clusters uncovered a 16-macrostate network with four highly-interconnected subnetworks consisting of four states each (Fig. 8C). The identities of the sixteen macrostates showed an exact correspondence to the sixteen possible *A/B* promoter binding configurations. This correspondence reflects the fact that, in the slow binding/unbinding, so-called non-adiabatic regime[71], the slow network dynamics are completely determined by unbinding and binding events that take the system from one promoter configuration macrostate to another, while all fluctuations in protein copy number occur on much faster timescales.

Each subnetwork in the MSM constructed with $C = 16$ corresponds to a single macrostate in the MSM constructed with $C = 4$. Thus, in the $C = 4$ MSM, four different promoter configurations are lumped together in a single macrostate, and dynamics of transitions among them is neglected. Counterintuitively, the locations of the $C = 4$ macrostates do not correspond directly to the four basins visible in the quasipotential landscape (Fig. 8B,D). Instead, the clusters combine distinct phenotypes—e.g., the red macrostate combines the *A/B* Lo/Lo and Lo/Hi phenotypes, because it includes the promoter configurations $A_{01} B_{10}$ and $A_{11} B_{10}$ (corresponding to Lo/Hi expression) and $A_{01} B_{00}$ and $A_{11} B_{00}$ (corresponding to Lo/Lo expression) (Fig. 8B, Table S3 and Fig. S4). This result demonstrates that the barriers visible in the quasipotential landscape do not reflect the slowest timescales in the system. This occurs because of the loss of information inherent to visualizing global dynamics via the quasipotential landscape, which often projects dynamics onto two system coordinates. In this case, projecting onto the protein *a* and protein *b* copy numbers loses information about the sixteen promoter configurations, obscuring the fact that barrier-crossing transitions can occur faster than some within-basin transitions. Plotting a time trajectory of brute force SSA simulations for this network supports the findings from the MSM: the dynamics shows frequent transitions within subnetworks, and less-frequent transitions between subnetworks, indicating the same hierarchy of system dynamics as was revealed by the 4- and 16-state MSMs (Fig. 8E).

**Transition path decomposition reveals nonequilibrium dynamics.** avMapping the most probable paths forward and backward between macrostate "1" (promoter configuration: $A_{01}B_{00}$) and macrostate "11" (promoter configuration: $A_{00}B_{01}$) revealed that a number of alternative transition paths are accessible to the network, and the paths typically transit between three and five intermediate macrostates. The decomposition shows three paths with significant (i.e., >15%) probability and 12 distinct paths with >1% probability (for both forward and backward transitions, Tables S3-4). The pathway decomposition also reveals a great deal of irreversibility in the forward and reverse transition paths, which is a hallmark of nonequilibrium dynamical systems[72]. For example, the most probable forward and reverse paths both transit three intermediates, but have only one intermediate (macrostate 5) in common (Fig. 8C and Table S4-5). Thus, the complete process of transitioning away from macrostate 1, through macrostate 11, and returning to 1 maps a dynamic cycle.

**DISCUSSION**

Our application of the MSM method to representative GRN motifs yielded dynamic insights with potential biological significance. Decomposition of transition pathways revealed that stochastic state-transitions between phenotypic states can occur via multiple alternative routes. Preference of the network to transition with higher likelihood through one particular pathway depended on the stability of intermediate macrostates, in a manner not directly intuitive from the steady-state probability landscape. The existence of "spurious attractors", or metastable intermediates that act as trap states to hinder stem cell reprogramming, has been discussed previously[11] as a general explanation for the existence of partially reprogrammed cells. By analogy, MSMs constructed in protein folding studies predict an ensemble of folding pathways, as well as the existence of misfolded trap states that reduce folding speed[56]. Our results suggest that multiple partially reprogrammed cell types could be accessible from a single initial cell state. Successful phenotype-transitions can occur predominantly through high-potential (unstable)—and thus difficult to observe experimentally—intermediate cell types. In future applications to specific gene GRNs, the MSM approach could predict a complex map of cell-reprogramming pathways, and thus potentially suggest combinations of targets towards improved safety and efficiency of reprogramming protocols.

Our study revealed that the two-gene MISA network can exhibit complex dynamic phenomena, involving a large number of metastable macrostates (up to 16), cycles and hierarchical dynamics, which can be conveniently visualized using the MSM. The quasipotential landscape has been used recently as a means of visualizing global dynamics and assessing locations and relative stabilities of phenotypic states of interest, in a manner that is quantitative (deriving strictly from underlying gene regulatory interactions), rather than qualitative or metaphorical (as was the case for the original Waddington epigenetic landscape)[21]. However, our study highlights the potential difficulty of interpreting global network dynamics based solely on the steady-state landscape, which is often projected onto one or two degrees of freedom. We found that phenotypically identical cell states—that is, network states marked by identical patterns of protein expression, inhabiting the same position in the projected landscape—can be separated by kinetic barriers, experiencing slow inter-conversion due to slow timescales for update to the epigenetic state (or promoter binding occupancy). Conversely, phenotypically distinct states marked by different levels of protein expression can be kinetically linked, experiencing relatively rapid inter-conversion. This type of stochastic inter-conversion is thought to occur in embryonic stem cells—for example, fluctuations in expression of the Nanog gene have been proposed to play a role in maintaining pluripotency[73], [74]. The hierarchical dynamics revealed by our study supports the idea that the phenotype of a cell could be more appropriately defined by dynamic patterns of regulator or marker expression levels[74], rather than on single-timepoint levels alone. This was seen in the 16-state MSM for the MISA network, where a given expression pattern (e.g., the Lo/Lo attractor) comprised multiple macrostates from separate dynamic subnetworks.

Complex, high-dimensional dynamical systems call for systematic methods of coarse-graining (or dimensionality reduction), for analysis of mechanisms and extraction of information that can be compared with experimental results. In the field of Molecular Dynamics, the complexity of, e.g., macromolecular conformational changes—involving thousands of atomic degrees of freedom and multiple dynamic intermediates—has driven the development of automated methods for prediction and analysis of essential system dynamics from simulations[75], [76]. In that field, coarse-graining has been achieved based on a variety of so-called geometric (structural) or, alternatively, kinetic clustering methods[50], [77]. Noe, et al.[77], discussed that geometric (or structure-based) coarse-graining methods can fail to produce an accurate description of system dynamics when structurally similar molecular conformations are separated by large energy barriers or, conversely, when dissimilar structures are connected by fast transitions, as they found in a study of polypeptide folding dynamics. In such cases, kinetic (i.e., separation-of-timescale-based) coarse-graining methods such as the MSM approach are more appropriate. Our application of the MSMs to GRNs demonstrates how similar complex dynamic phenomena can manifest at the "network"-scale.

The challenge of solving the CME due to the curse-of-dimensionality is well known. The MSM approach is related to other projection-based model reduction methods that aim to reduce the computational burden of solving the CME directly by projecting the rate (or transition) matrix onto a smaller subspace or aggregated state-space with fewer degrees of freedom. Such approaches include the Finite State Projection algorithm[40], and methods based on Krylov subspaces[28], [78], [79], sparse-gridding[80], and separation-of-timescales[29], [81], [82]. The MSM is distinct from other timescale-based approaches in that, rather than partitioning the system into categories of slow versus fast reactions[81] or species[29], or basing categories on physical intuition[82], it systematically groups microstates in such a way that maximizes metastability of aggregated states[62]. The practical benefit of this approach is its capacity to describe a system compactly in terms of long-lived, perhaps experimentally observable, states. Another important distinction between the MSM approach and other CME model reduction methods is that its primary end-goal is *not* to solve the CME per se. Rather, the emphasis in studies employing MSMs has generally been on gaining mechanistic, physical, or experimentally-relevant insights to complex system dynamics[83]–[85]. As such, the approach does not optimally balance the tradeoff between computational expense versus quantitative accuracy of the solution, as other methods have done explicitly[86]. Instead, the method can be considered to balance the tradeoff between accuracy and "human-interpretability", where decreasing the number of macrostates preserved in the MSM coarse-graining tends to favor the latter over the former.

A potential drawback of the workflow presented in this paper is that it requires an enumeration of the system state-space in order to construct the biochemical rate matrix **K**. Networks of increased complexity or molecular copy numbers will lead to prohibitively large matrix sizes. Here, we restricted our study to model systems with a relatively small number of reachable microstates (i.e., $\sim 10^4$ microstates permitted tractable computations on desktop computers with MATLAB[43]). However, an advantage of the MSM approach is its use of the stochastic transition matrix $\mathbf{T}(\tau)$ (rather than **K**), which can be estimated from simulations by sampling transition counts between designated regions of state-space in trajectories of length $\tau$ [87]. Systems of increased complexity/dimensionality are generally more accessible to simulations, because the size of the state-space is automatically restricted to those states visited within finite-length simulations. In ongoing work (Tse, *et al.*, in preparation), we find that the MSM approach interfaces well with SSA simulations of biochemical network dynamics, combined with enhanced sampling techniques[88], [89]. We anticipate that, as in the Molecular Dynamics field, the MSM framework in applications to biochemical networks will prove useful as a tool for post-processing simulation data. Furthermore, we anticipate that the approach could potentially interface with other numerical approximation techniques that have been developed in recent years for reduction of the CME.

A potential challenge for the application of the PCCA+-based spectral clustering method to biochemical networks is that, as open systems, biochemical networks generally do not obey detailed balance. This means that the stochastic transition matrices do not have the property of irreversibility, which was originally taken to be a requirement for application of the PCCA algorithm[50]. However, later work by Roblitz et al.[48], found that the PCCA+ method also delivers an optimal clustering for irreversible systems. In this study, we found that the PCCA+ method could determine appropriate clusters in GRNs, and could furthermore uncover nonequilibrium cycles, as seen in the irreversibility (distinct forward and backward) of transition paths in the 16-state system. Newer methods of MSM building, which are specifically designed to treat nonequilibrium dynamical systems, have appeared recently[90]. It may prove fruitful to explore these alternative methods in order to identify the most appropriate, general MSM framework for application to various biochemical networks.

**CONCLUSIONS**

In this work, we present a method for analyzing multistability and global state-switching dynamics in gene networks modeled by stochastic chemical kinetics, using the MSM framework. We found that the approach is able to: (1) identify the number and identities of long-lived phenotypic-states, or network "macrostates", (2) predict the steady-state probabilities of all macrostates along with probabilities of transitioning to other macrostates on a given timescale, and (3) decompose global dynamics into a set of dominant transition pathways and their associated relative probabilities, linking two system states of interest. Because the method is based on the discrete-space, stochastic transition matrix, it correctly identified stochastic multistability where a continuum model failed to find multiple steady states. The quantitative accuracy of the dynamics propagated by the coarse-grained MSM was highest in a parameter regime with slow DNA-binding and unbinding kinetics, indicating that in GRNs the assumption of memory-less hopping among a small number of macrostates is most valid in this regime. By projecting dynamics encompassing a large state-space onto a tractable number of macrostates, the MSMs revealed complex dynamic phenomena in GRNs, including hierarchical dynamics, nonequilibrium cycles, and alternative possible routes for phenotypic state-transitions. The ability to unravel these processes using the MSM framework can shed light on regulatory mechanisms that govern cell phenotype stability, and inform experimental reprogramming strategies. The MSM provides an intuitive representation of complex biological dynamics operating over multiple timescales, which in turn can provide the key to decoding biological mechanisms. Overall, our results demonstrate that the MSM framework—which has been generally applied thus far in the context of molecular dynamics via atomistic simulations—can be a useful tool for visualization and analysis of complex, multistable dynamics in gene networks, and in biochemical reaction networks more generally.

**LIST OF ABBREVIATIONS**

GRNs – Gene Regulatory Networks

GRN – Gene Regulatory Network

CME – Chemical Master Equation

SSA – Stochastic Simulation Algorithm

MSM – Markov State Model

ETS - Exclusive Toggle Switch

MISA – Mutual Inhibition/Self-Activation

PCCA+- Robust Perron Cluster Analysis

ODE – Ordinary Differential Equation

# DECLARATIONS

**Ethics approval and consent to participate**

Not applicable.

**Consent for publication**

Not applicable.

**Availability of data and materials**

The datasets supporting the conclusions of this article are included in the Additional File 1.

**Competing interests**

The authors declare that they have no competing interests.

**Funding**

We acknowledge financial support from the UC Irvine Henry Samueli School of Engineering.

**Authors' contributions**

BC and ER designed and performed research. MT contributed to data analysis and manuscript preparation. RS contributed to data analysis. BC and ER wrote the manuscript. All authors read and approved the final manuscript.

**Acknowledgements**

We thank Jun Allard helpful discussions.

**References**

[1] A. Arkin, J. Ross, and H. McAdams, "Stochastic kinetic analysis of developmental pathway bifurcation in phage lambda-infected Escherichia coli cells," *Genetics*, vol. 149, no. 75232100002, pp. 1633–1648, Dec. 1997.

[2] W. Xiong and J. E. Ferrell, "A positive-feedback-based bistable 'memory module' that governs a cell fate decision," *Nature*, vol. 426, no. 6965, pp. 460–465, Nov. 2003.

[3] J. Zhou and S. Huang, "Understanding gene circuits at cell-fate branch points for rational cell reprogramming," *Trends Genet. TIG*, vol. 27, no. 21146896, pp. 55–62, Feb. 2011.

[4] M. Lu, M. K. Jolly, R. Gomoto, B. Huang, J. Onuchic, and E. Ben-Jacob, "Tristability in Cancer-Associated MicroRNA-TF Chimera Toggle Switch," *J. Phys. Chem. B*, vol. 117, no. 42, pp. 13164–13174, Oct. 2013.

[5] H. Feng and J. Wang, "A new mechanism of stem cell differentiation through slow binding/unbinding of regulators to genes," *Sci. Rep.*, vol. 2, no. 22870379, p. 550, Jan. 2012.

[6] B. Zhang and P. Wolynes, "Stem cell differentiation as a many-body problem," *Proc. Natl. Acad. Sci. U. S. A.*, vol. 111, no. 24946805, pp. 10185–90, Jul. 2014.

[7] P. Wang, C. Song, H. Zhang, Z. Wu, X.-J. Tian, and J. Xing, "Epigenetic state network approach for describing cell phenotypic transitions," *Interface Focus*, vol. 4, no. 3, p. 20130068, Jun. 2014.

[8] T. Hong, J. Xing, L. Li, and J. Tyson, "A mathematical model for the reciprocal differentiation of T helper 17 cells and induced regulatory T cells," *PLoS Comput. Biol.*, vol. 7, no. 21829337, p. e1002122, Jul. 2011.

[9] T. Graf and T. Enver, "Forcing cells to change lineages," *Nature*, vol. 462, no. 19956253, pp. 587–94, Dec. 2009.

[10] S. Huang, "The molecular and mathematical basis of Waddington's epigenetic landscape: a framework for post-Darwinian biology?," *BioEssays News Rev. Mol. Cell. Dev. Biol.*, vol. 34, no. 22102361, pp. 149–57, Feb. 2012.

[11] A. H. Lang, H. Li, J. J. Collins, and P. Mehta, "Epigenetic Landscapes Explain Partially Reprogrammed Cells and Identify Key Reprogramming Genes," *PLoS Comput. Biol.*, vol. 10, no. 8, p. e1003734, Aug. 2014.

[12] M. B. Elowitz, "Stochastic Gene Expression in a Single Cell," *Science*, vol. 297, no. 5584, pp. 1183–1186, Aug. 2002.

[13] E. M. Ozbudak, M. Thattai, I. Kurtser, A. D. Grossman, and A. van Oudenaarden, "Regulation of noise in the expression of a single gene," *Nat. Genet.*, vol. 31, no. 1, pp. 69–73, May 2002.

[14] I. Golding, J. Paulsson, S. M. Zawilski, and E. C. Cox, "Real-Time Kinetics of Gene Activity in Individual Bacteria," *Cell*, vol. 123, no. 6, pp. 1025–1036, Dec. 2005.

[15] N. Balaban, J. Merrin, R. Chait, L. Kowalik, and S. Leibler, "Bacterial persistence as a phenotypic switch," *Science*, vol. 305, no. 15308767, pp. 1622–5, Sep. 2004.

[16] M. Acar, J. T. Mettetal, and A. van Oudenaarden, "Stochastic switching as a survival strategy in fluctuating environments," *Nat. Genet.*, vol. 40, no. 4, pp. 471–475, Apr. 2008.

[17] S. Sharma, D. Lee, B. Li, M. Quinlan, F. Takahashi, S. Maheswaran, U. McDermott, N. Azizian, L. Zou, M. Fischbach, K. Wong, K. Brandstetter, B. Wittner, S. Ramaswamy, M. Classon, and J. Settleman, "A chromatin-mediated reversible

drug-tolerant state in cancer cell subpopulations," *Cell*, vol. 141, no. 20371346, pp. 69–80, Apr. 2010.

[18] H. H. Chang, M. Hemberg, M. Barahona, D. E. Ingber, and S. Huang, "Transcriptome-wide noise controls lineage choice in mammalian progenitor cells," *Nature*, vol. 453, no. 7194, pp. 544–547, May 2008.

[19] J.-E. Dietrich and T. Hiiragi, "Stochastic patterning in the mouse pre-implantation embryo," *Dev. Camb. Engl.*, vol. 134, no. 23, pp. 4219–4231, Dec. 2007.

[20] L. Yuan, G. C. Chan, D. Beeler, L. Janes, K. C. Spokes, H. Dharaneeswaran, A. Mojiri, W. J. Adams, T. Sciuto, G. Garcia-Cardeña, G. Molema, P. M. Kang, N. Jahroudi, P. A. Marsden, A. Dvorak, E. R. Regan, and W. C. Aird, "A role of stochastic phenotype switching in generating mosaic endothelial cell heterogeneity," *Nat. Commun.*, vol. 7, p. 10160, 2016.

[21] C. Waddington and H. Kacser, *The Strategy of the Genes*. Routledge, 1957.

[22] J. Wang, K. Zhang, L. Xu, and E. Wang, "Quantifying the Waddington landscape and biological paths for development and differentiation," *Proc. Natl. Acad. Sci. U. S. A.*, vol. 108, no. 21536909, pp. 8257–62, May 2011.

[23] G. Karlebach and R. Shamir, "Modelling and analysis of gene regulatory networks," *Nat. Rev. Mol. Cell Biol.*, vol. 9, no. 10, pp. 770–780, Oct. 2008.

[24] T. Kepler and T. Elston, "Stochasticity in transcriptional regulation: Origins, consequences, and mathematical representations," *Biophys. J.*, vol. 81, no. 172407800010, pp. 3116–3136, Dec. 2000.

[25] D. GILLESPIE, "EXACT STOCHASTIC SIMULATION OF COUPLED CHEMICAL-REACTIONS," *J. Phys. Chem.*, vol. 81, no. A1977EE49800008, pp. 2340–2361, Dec. 1976.

[26] B. Munsky and M. Khammash, "The finite state projection algorithm for the solution of the chemical master equation," *J. Chem. Phys.*, vol. 124, no. 16460146, p. 44104, Jan. 2006.

[27] Y. Cao and J. Liang, "Optimal enumeration of state space of finitely buffered stochastic molecular networks and exact computation of steady state landscape probability," *BMC Syst. Biol.*, vol. 2, no. 1, p. 30, 2008.

[28] V. Wolf, R. Goel, M. Mateescu, and T. A. Henzinger, "Solving the chemical master equation using sliding windows," *BMC Syst. Biol.*, vol. 4, no. 1, p. 42, 2010.

[29] C. D. Pahlajani, P. J. Atzberger, and M. Khammash, "Stochastic reduction method for biological chemical kinetics using time-scale separation," *J. Theor. Biol.*, vol. 272, no. 1, pp. 96–112, Mar. 2011.

[30] R. B. Sidje and H. D. Vo, "Solving the chemical master equation by a fast adaptive finite state projection based on the stochastic simulation algorithm," *Math. Biosci.*, vol. 269, pp. 10–16, Nov. 2015.

[31] S. Huang, Y.-P. Guo, G. May, and T. Enver, "Bifurcation dynamics in lineage-commitment in bipotent progenitor cells," *Dev. Biol.*, vol. 305, no. 2, pp. 695–713, May 2007.

[32] R. Ma, J. Wang, Z. Hou, and H. Liu, "Small-number effects: a third stable state in a genetic bistable toggle switch," *Phys. Rev. Lett.*, vol. 109, no. 23368390, p. 248107, Dec. 2012.

[33] Y. Cao, H.-M. Lu, and J. Liang, "Probability landscape of heritable and robust epigenetic state of lysogeny in phage lambda," *Proc. Natl. Acad. Sci.*, vol. 107, no. 43, pp. 18445–18450, Oct. 2010.

[34] B. Munsky, Z. Fox, and G. Neuert, "Integrating single-molecule experiments and discrete stochastic models to understand heterogeneous gene transcription dynamics," *Methods*, vol. 85, pp. 12–21, Sep. 2015.

[35] C. Schütte, A. Fischer, W. Huisinga, and P. Deuflhard, "A Direct Approach to Conformational Dynamics Based on Hybrid Monte Carlo," *J. Comput. Phys.*, vol. 151, no. 1, pp. 146–168, May 1999.

[36] W. C. Swope, J. W. Pitera, and F. Suits, "Describing Protein Folding Kinetics by Molecular Dynamics Simulations. 1. Theory †," *J. Phys. Chem. B*, vol. 108, no. 21, pp. 6571–6581, May 2004.

[37] D. S. Chekmarev, T. Ishida, and R. M. Levy, "Long-Time Conformational Transitions of Alanine Dipeptide in Aqueous Solution: Continuous and Discrete-State Kinetic Models," *J. Phys. Chem. B*, vol. 108, no. 50, pp. 19487–19495, Dec. 2004.

[38] G. R. Bowman, X. Huang, and V. S. Pande, "Network models for molecular kinetics and their initial applications to human health," *Cell Res.*, vol. 20, no. 6, pp. 622–630, Jun. 2010.

[39] P. Érdi and J. Tóth, *Mathematical models of chemical reactions: theory and applications of deterministic and stochastic models*. Princeton, N.J: Princeton University Press, 1989.

[40] B. Munsky and M. Khammash, "The finite state projection algorithm for the solution of the chemical master equation," *J. Chem. Phys.*, vol. 124, no. 4, p. 44104, 2006.

[41] K. R. Sanft, S. Wu, M. Roh, J. Fu, R. K. Lim, and L. R. Petzold, "StochKit2: software for discrete stochastic simulation of biochemical systems with events," *Bioinformatics*, vol. 27, no. 17, pp. 2457–2458, Sep. 2011.

[42] N. G. van Kampen, *Stochastic processes in physics and chemistry*. Amsterdam; Boston; London: Elsevier, 2007.

[43] *MATLAB Release 2015a.* Natick, Massachusetts, United States: The MathWorks, Inc.

[44] M. K. Scherer, B. Trendelkamp-Schroer, F. Paul, G. Pérez-Hernández, M. Hoffmann, N. Plattner, C. Wehmeyer, J.-H. Prinz, and F. Noé, "PyEMMA 2: A Software Package for Estimation, Validation, and Analysis of Markov Models," *J. Chem. Theory Comput.*, vol. 11, no. 11, pp. 5525–5542, Nov. 2015.

[45] J.-H. Prinz, H. Wu, M. Sarich, B. Keller, M. Senne, M. Held, J. D. Chodera, C. Schütte, and F. Noé, "Markov models of molecular kinetics: Generation and validation," *J. Chem. Phys.*, vol. 134, no. 17, p. 174105, 2011.

[46] J. D. Chodera and F. Noé, "Markov state models of biomolecular conformational dynamics," *Curr. Opin. Struct. Biol.*, vol. 25, pp. 135–144, Apr. 2014.

[47] P. Deuflhard, W. Huisinga, A. Fischer, and C. Schütte, "Identification of almost invariant aggregates in reversible nearly uncoupled Markov chains," *Linear Algebra Its Appl.*, vol. 315, no. 1–3, pp. 39–59, Aug. 2000.

[48] S. Röblitz and M. Weber, "Fuzzy spectral clustering by PCCA+: application to Markov state models and data classification," *Adv. Data Anal. Classif.*, vol. 7, no. 2, pp. 147–179, Jun. 2013.

[49] P. Deuflhard, W. Huisinga, A. Fischer, and C. Schütte, "Identification of almost invariant aggregates in reversible nearly uncoupled Markov chains," *Linear Algebra Its Appl.*, vol. 315, no. 1–3, pp. 39–59, Aug. 2000.

[50] P. Deuflhard and M. Weber, "Robust Perron cluster analysis in conformation dynamics," *Linear Algebra Its Appl.*, vol. 398, pp. 161–184, Mar. 2005.

[51] N.-V. Buchete and G. Hummer, "Coarse Master Equations for Peptide Folding Dynamics †," *J. Phys. Chem. B*, vol. 112, no. 19, pp. 6057–6069, May 2008.

[52] K. A. Beauchamp, G. R. Bowman, T. J. Lane, L. Maibaum, I. S. Haque, and V. S. Pande, "MSMBuilder2: Modeling Conformational Dynamics on the Picosecond to Millisecond Scale," *J. Chem. Theory Comput.*, vol. 7, no. 10, pp. 3412–3419, Oct. 2011.

[53] S. Kube and M. Weber, "A coarse graining method for the identification of transition rates between molecular conformations," *J. Chem. Phys.*, vol. 126, no. 2, p. 24103, 2007.

[54] W. E. and E. Vanden-Eijnden, "Towards a Theory of Transition Paths," *J. Stat. Phys.*, vol. 123, no. 3, pp. 503–523, May 2006.

[55] P. Metzner, C. Schütte, and E. Vanden-Eijnden, "Transition Path Theory for Markov Jump Processes," *Multiscale Model. Simul.*, vol. 7, no. 3, pp. 1192–1219, Jan. 2009.

[56] F. Noe, C. Schutte, E. Vanden-Eijnden, L. Reich, and T. R. Weikl, "Constructing the equilibrium ensemble of folding pathways from short off-equilibrium simulations," *Proc. Natl. Acad. Sci.*, vol. 106, no. 45, pp. 19011–19016, Nov. 2009.

[57] D. Schultz, A. M. Walczak, J. N. Onuchic, and P. G. Wolynes, "Extinction and resurrection in gene networks," *Proc. Natl. Acad. Sci.*, vol. 105, no. 49, pp. 19165–19170, Dec. 2008.

[58] H. Feng and J. Wang, "A new mechanism of stem cell differentiation through slow binding/unbinding of regulators to genes," *Sci. Rep.*, vol. 2, Aug. 2012.

[59] M. J. Morelli, S. Tănase-Nicola, R. J. Allen, and P. R. ten Wolde, "Reaction Coordinates for the Flipping of Genetic Switches," *Biophys. J.*, vol. 94, no. 9, pp. 3413–3423, May 2008.

[60] S. Huang, "Reprogramming cell fates: reconciling rarity with robustness," *BioEssays News Rev. Mol. Cell. Dev. Biol.*, vol. 31, no. 19319911, pp. 546–60, May 2009.

[61] S. Huang, "Hybrid T-Helper Cells: Stabilizing the Moderate Center in a Polarized System," *PLoS Biol.*, vol. 11, no. 8, p. e1001632, Aug. 2013.

[62] V. S. Pande, K. Beauchamp, and G. R. Bowman, "Everything you wanted to know about Markov State Models but were afraid to ask," *Methods*, vol. 52, no. 1, pp. 99–105, Sep. 2010.

[63] Gardner, Timothy, S. Charles, R. Cantor, and James J. Collins, "Construction of a genetic toggle switch in Escherichia coli," *Nature*, vol. 403, no. 6767, pp. 339–342, 2000.

[64] A. Lipshtat, A. Loinger, N. Balaban, and O. Biham, "Genetic toggle switch without cooperative binding," *Phys. Rev. Lett.*, vol. 96, no. 16712399, p. 188101, May 2006.
[65] D. Schultz, A. M. Walczak, J. N. Onuchic, and P. G. Wolynes, "Extinction and resurrection in gene networks," *Proc. Natl. Acad. Sci. U. S. A.*, vol. 105, no. 49, pp. 19165–19170, Dec. 2008.
[66] T. J. Lane, G. R. Bowman, K. Beauchamp, V. A. Voelz, and V. S. Pande, "Markov State Model Reveals Folding and Functional Dynamics in Ultra-Long MD Trajectories," *J. Am. Chem. Soc.*, vol. 133, no. 45, pp. 18413–18419, Nov. 2011.
[67] M. J. Tse, B. K. Chu, M. Roy, and E. L. Read, "DNA-Binding Kinetics Determines the Mechanism of Noise-Induced Switching in Gene Networks," *Biophys. J.*, vol. 109, no. 8, pp. 1746–1757, Oct. 2015.
[68] W. E. and E. Vanden-Eijnden, "Towards a Theory of Transition Paths," *J. Stat. Phys.*, vol. 123, no. 3, pp. 503–523, May 2006.
[69] P. Metzner, C. Schütte, and E. Vanden-Eijnden, "Transition Path Theory for Markov Jump Processes," *Multiscale Model. Simul.*, vol. 7, no. 3, pp. 1192–1219, Jan. 2009.
[70] A. Berezhkovskii, G. Hummer, and A. Szabo, "Reactive flux and folding pathways in network models of coarse-grained protein dynamics," *J. Chem. Phys.*, vol. 130, no. 20, p. 205102, 2009.
[71] A. M. Walczak, J. N. Onuchic, and P. G. Wolynes, "Absolute rate theories of epigenetic stability," *Proc. Natl. Acad. Sci.*, vol. 102, no. 52, pp. 18926–18931, Dec. 2005.
[72] J. Wang, K. Zhang, and E. Wang, "Kinetic paths, time scale, and underlying landscapes: A path integral framework to study global natures of nonequilibrium systems and networks," *J. Chem. Phys.*, vol. 133, no. 12, p. 125103, 2010.
[73] I. Chambers, J. Silva, D. Colby, J. Nichols, B. Nijmeijer, M. Robertson, J. Vrana, K. Jones, L. Grotewold, and A. Smith, "Nanog safeguards pluripotency and mediates germline development," *Nature*, vol. 450, no. 7173, pp. 1230–1234, Dec. 2007.
[74] T. Kalmar, C. Lim, P. Hayward, S. Muñoz-Descalzo, J. Nichols, J. Garcia-Ojalvo, and A. Martinez Arias, "Regulated Fluctuations in Nanog Expression Mediate Cell Fate Decisions in Embryonic Stem Cells," *PLoS Biol.*, vol. 7, no. 7, p. e1000149, Jul. 2009.
[75] J. D. Chodera, N. Singhal, V. S. Pande, K. A. Dill, and W. C. Swope, "Automatic discovery of metastable states for the construction of Markov models of macromolecular conformational dynamics," *J. Chem. Phys.*, vol. 126, no. 15, p. 155101, 2007.
[76] G. R. Bowman, K. A. Beauchamp, G. Boxer, and V. S. Pande, "Progress and challenges in the automated construction of Markov state models for full protein systems," *J. Chem. Phys.*, vol. 131, no. 12, p. 124101, 2009.
[77] G. Pérez-Hernández, F. Paul, T. Giorgino, G. De Fabritiis, and F. Noé, "Identification of slow molecular order parameters for Markov model construction," *J. Chem. Phys.*, vol. 139, no. 1, p. 15102, 2013.

[78] Burrage, Kevin, Hegland, M., Macnamara, Shev, & Sidje, Roger, "A Krylov-based finite state projection algorithm for solving the chemical master equation arising in the discrete modelling of biological systems," in *Proceedings of the Markov 150th Anniversary Conference*, 2006.

[79] Y. Cao, A. Terebus, and J. Liang, "Accurate Chemical Master Equation Solution Using Multi-Finite Buffers," *Multiscale Model. Simul.*, vol. 14, no. 2, pp. 923–963, Jan. 2016.

[80] M. Hegland, C. Burden, L. Santoso, S. MacNamara, and H. Booth, "A solver for the stochastic master equation applied to gene regulatory networks," *J. Comput. Appl. Math.*, vol. 205, no. 2, pp. 708–724, Aug. 2007.

[81] E. L. Haseltine and J. B. Rawlings, "Approximate simulation of coupled fast and slow reactions for stochastic chemical kinetics," *J. Chem. Phys.*, vol. 117, no. 15, p. 6959, 2002.

[82] S. Peleš, B. Munsky, and M. Khammash, "Reduction and solution of the chemical master equation using time scale separation and finite state projection," *J. Chem. Phys.*, vol. 125, no. 20, p. 204104, 2006.

[83] Y. Kuroda, A. Suenaga, Y. Sato, S. Kosuda, and M. Taiji, "All-atom molecular dynamics analysis of multi-peptide systems reproduces peptide solubility in line with experimental observations," *Sci. Rep.*, vol. 6, p. 19479, Jan. 2016.

[84] G. Jayachandran, V. Vishal, and V. S. Pande, "Using massively parallel simulation and Markovian models to study protein folding: Examining the dynamics of the villin headpiece," *J. Chem. Phys.*, vol. 124, no. 16, p. 164902, 2006.

[85] N. Singhal, C. D. Snow, and V. S. Pande, "Using path sampling to build better Markovian state models: Predicting the folding rate and mechanism of a tryptophan zipper beta hairpin," *J. Chem. Phys.*, vol. 121, no. 1, p. 415, 2004.

[86] J. J. Tapia, J. R. Faeder, and B. Munsky, "Adaptive coarse-graining for transient and quasi-equilibrium analyses of stochastic gene regulation," 2012, pp. 5361–5366.

[87] J. L. Adelman and M. Grabe, "Simulating rare events using a weighted ensemble-based string method," *J. Chem. Phys.*, vol. 138, no. 4, p. 44105, 2013.

[88] D. Bhatt and I. Bahar, "An adaptive weighted ensemble procedure for efficient computation of free energies and first passage rates," *J. Chem. Phys.*, vol. 137, no. 10, p. 104101, Sep. 2012.

[89] B. W. Zhang, D. Jasnow, and D. M. Zuckerman, "The 'weighted ensemble' path sampling method is statistically exact for a broad class of stochastic processes and binning procedures," *J. Chem. Phys.*, vol. 132, no. 5, p. 54107, 2010.

[90] Weber, Marcus, and K. Fackeldey, "G-pcca: Spectral clustering for non-reversible markov chains," *ZIB Rep.*, vol. 15, no. 35.

**Figure Titles and Legends**

Figure 1.

Eigenvalue and eigenvector analysis of the Mutual Inhibition/Self Activation (MISA) network. (A) Schematic of the MISA network motif. (B) The fifteen largest eigenvalues of the stochastic transition matrix $\mathbf{T}(\tau)$, indexed in descending order, for $\tau = 5$ (circles) and $\tau = 0.5$ (crosses) (time units of inverse protein degradation rate, $k^{-1}$). Gaps indicate separation between processes occurring on different timescales. Network parameter values are listed in Table S1. (C) The quasipotential landscape (left) and probability landscape (right) for the MISA motif, projected onto the A vs. B protein copy number subspace, showing four visible attractors. Landscapes were obtained from $\phi_1$, the eigenvector associated with the largest eigenvalue of $\mathbf{T}(\tau)$. (D) Left to right: second, third, and fourth eigenvectors ($\phi_2$, $\phi_3$, $\phi_4$) of $\mathbf{T}(\tau)$. The sign structure reveals the nature of the slowest dynamical processes (see text).

Figure 2.

Four metastable clusters, or network "macrostates", identified for the MISA network by the Markov State Model approach. (Rate parameters same as Fig. 1) (A) Macrostate centers located by their respective 50% probability contours, corresponding to visible peaks in the probability landscape. B) Schematics of the most probable gene promoter configuration for each metastable cluster.

Figure 3.

Dependence of the MISA network eigenvalues, landscape, and MSM on the repressor unbinding parameter $f_r$. Top to Bottom: increasing $fr = \{1E-3, 1E-2, 1E-1, 1\}$ in units of protein degradation rate, $k^{-1}$ (complete parameter list in Table S1). (A) The eigenvalue spectrum of $\mathbf{T}(\tau)$ for $\tau = 5$, and associated timescales. (B) The quasipotential landscape. (C) The Markov State Model with four macrostates, visualized by the 50% probability contour for each metastable state. (D) The state transition graph. Nodes and edges denote macrostates and transition probabilities, respectively. The size of each node is proportional to the steady-state probability, and edge thickness is proportional to the probability of transition within $\tau = 5$.

Figure 4.

Comparison of ODE and MSM analysis of the monomer Exclusive Toggle Switch (ETS) network. (A) Schematic of the ETS network motif. (B) The Markov State Model identifies three macrostates corresponding to the Hi/Lo, Hi/Hi, and Lo/Hi phenotypes. Parameter values are listed in Table S2. (C) The nulllclines and vector field of the deterministic ODEs show a single fixed point steady-state, with both genes expressing at the maximum rate (Hi/Hi phenotype). (B,D,E) The corresponding landscape and MSM show tristability: (D) The quasipotential landscape shows three visible attractors corresponding to the Hi/Lo, Hi/Hi, and Lo/Hi phenotypes. (E) The 20 dominant eigenvalues reveal timescale separation, including a gap after $\lambda_3$.

Figure 5.

MSM approximation error for the MISA motif. Relaxation of the system from a particular initial configuration (see text), as obtained from (A) the full transition matrix, (B) brute force SSA simulation, and (C) the reduced transition matrix obtained from the MSM. Color-coding is according to the macrostates, as in Figs. 1-3: blue, black, red, green correspond to A/B expression phenotypes Hi/Lo, Hi/Hi, Lo/Hi, and Lo/Lo, respectively. (D) Calculated approximation error as a function of time, comparing the reduced MSM to the full CME dynamics. Network parameter values are same as Figs. 1,2.

Figure 6.

The MSM approximation accuracy for the ETS motif depends on rate parameters and number of macrostates in the reduced model. (A) Quasipotential landscape for the exclusive dimeric repressor toggle switch, with increasing DNA-binding rates (left to right: $fr = \{1E-4, 1E-2, 1E0, 1E2, 1E4\}$, all parameter values listed in Table S2), demonstrating the dependence of basin number and barrier height on network parameters. (B) Global error of the MSM approximation. Left: Global error as a function of time (in intervals of $\tau$) for different $f_r$ and numbers of macrostates. Solid lines: global error of the 3-state MSM. Dashed lines: global error of the 2-state MSM. Right: Total global error over k$\tau$, $k = 0$ to 500, for a 3-state (solid blue) or 2-state (dashed blue) MSM. Solid orange line: the longest system lifetime $t_2$. (C) Error of the MSM approximation when the system is initialized in a particular microstate. Left: Error as a function of time (in intervals of $\tau$) for different adiabaticities and different numbers of macrostates. Solid lines: error of the 3-state MSM. Dashed lines: error of the 2-state MSM. Right: Total error from a particular microstate over k$\tau$ where $k = 0$ to 500, for a 3-state (solid blue) or 2-state (dashed blue) MSM. Orange line: the longest system lifetime $t_2$.

Figure 7.

Dependence of stochastic transition paths on the repressor unbinding rate parameter $f_r$ in the MISA network (parameter values listed in Table S1). (A)Table of all possible transition paths starting from the Hi/Lo (blue) and ending in the Lo/Hi (red) macrostate (color coding is same as Figs. 1-3 and Fig. 5). Relative probabilities of traversing a given path are shown, along with the stationary probabilities of the system to be found in a given macrostate. (B-D) Dominant transition paths superimposed on the 3D quasipotential surfaces for $fr = \{5E-4, 1E-3, 5E-3\}$, demonstrating how dominant paths can traverse high-potential areas of the landscape. For example, when $f_r = 1E-3$, (panel C), successful transitions most likely go through the Hi/Hi state (3.2% populated at steady state), though this requires a large barrier crossing. Pathway percentages are superimposed on the landscapes.

Figure 8:

Hierarchical dynamics revealed by MSM analysis of the MISA network in the slow DNA-binding/unbinding parameter regime. All network parameters listed in Table S1. (A) Eigenvalue spectrum of $\mathbf{T}(\tau)$, $\tau =1$, showing 16 dominant eigenvalues. (B) 4-macrostate MSM: 70% probability contours superimposed onto the quasipotential surface. In this parameter regime, separate attractors in the landscape are kinetically linked in the same subnetwork (see text). (C) 16-macrostate MSM showing 4 highly connected subnetworks (colored ovals). Each macrostate corresponds to a particular promoter binding-configuration (see numbering scheme in Table S5). A pair of representative transition paths through the network are highlighted. Red path: most probable forward transition path from macrostate 1 to macrostate 11. Blue path: most probable reverse path from 11 to 1. (D) State transition graph for the 4-macrostate MSM. (E) Brute force SSA simulation of the MISA network over time. Trajectory is plotted according to the 16-macrostate (promoter configuration) indexing as in panel C and Table S5. Colored panels reflect the four subnetworks/$C = 4$ macrostates. Orange inset: zoomed in trajectory segment, showing a switching event between the red and green subnetworks.

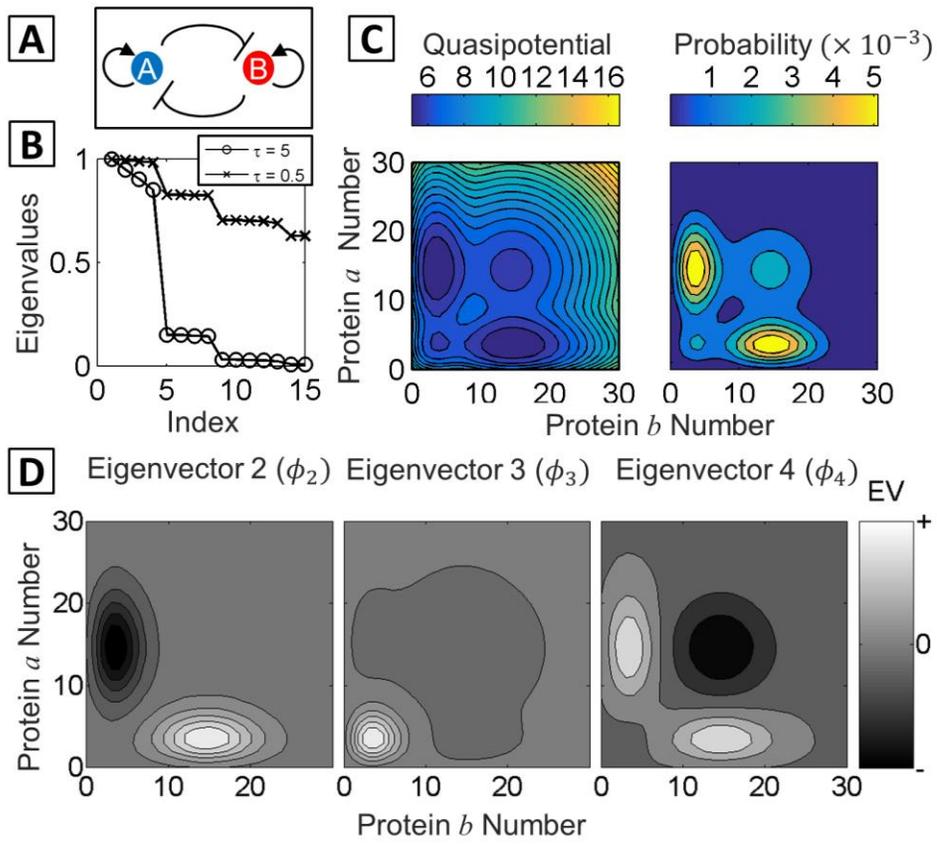

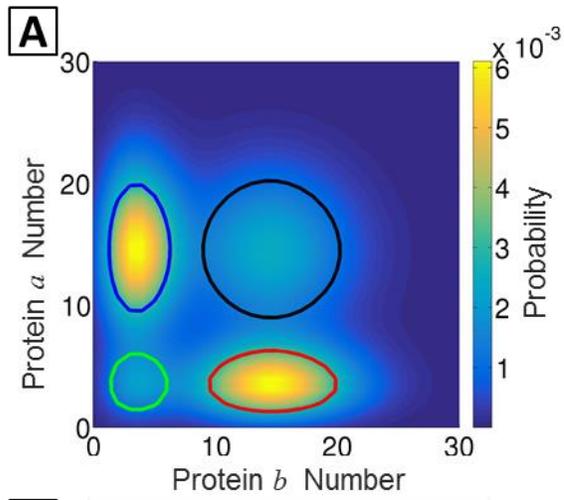
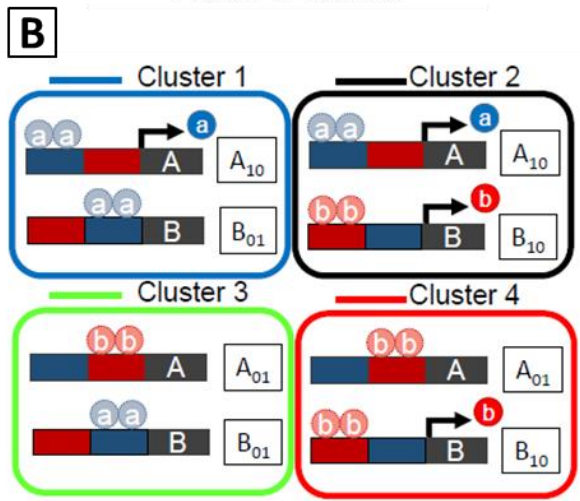

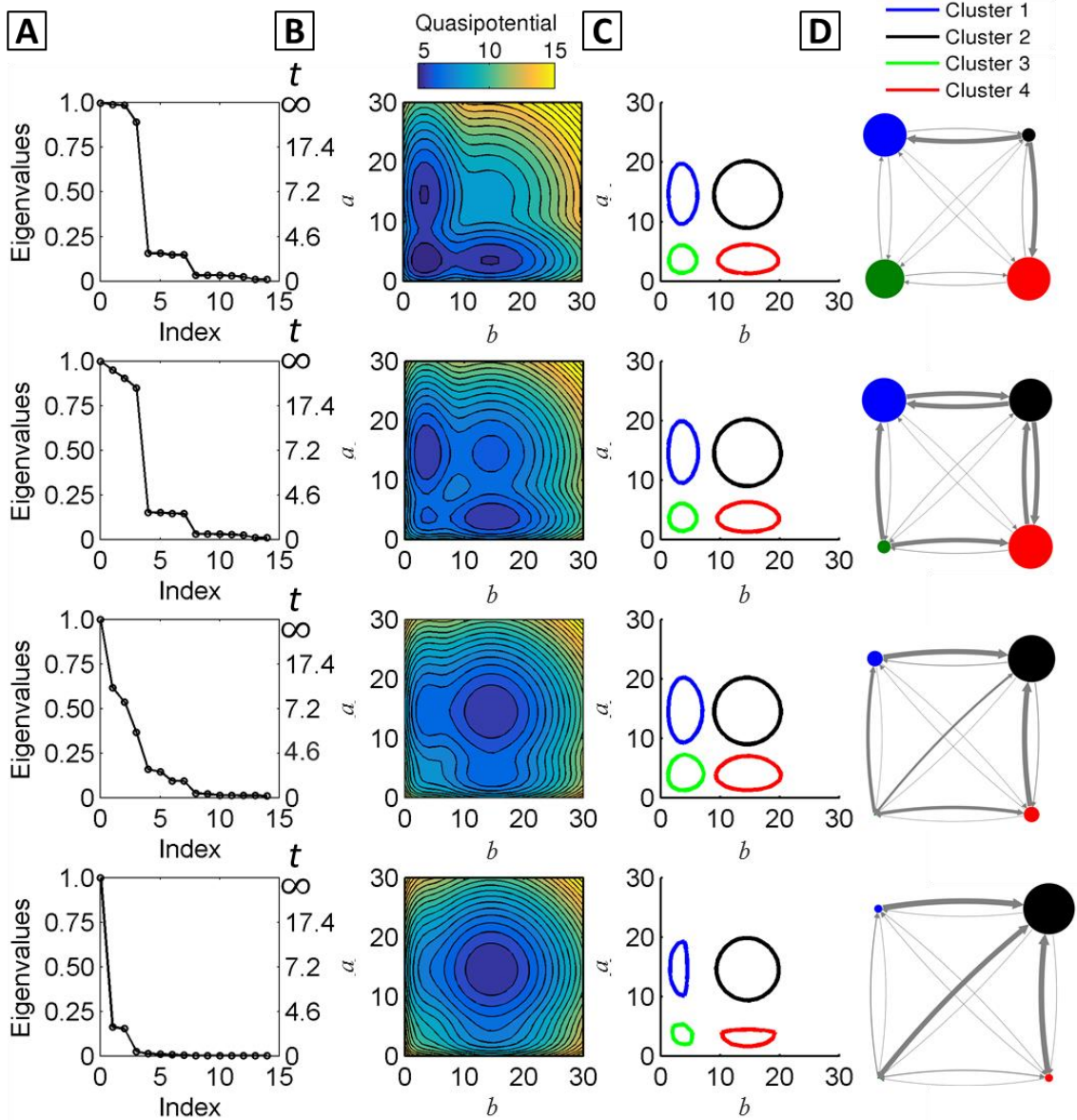

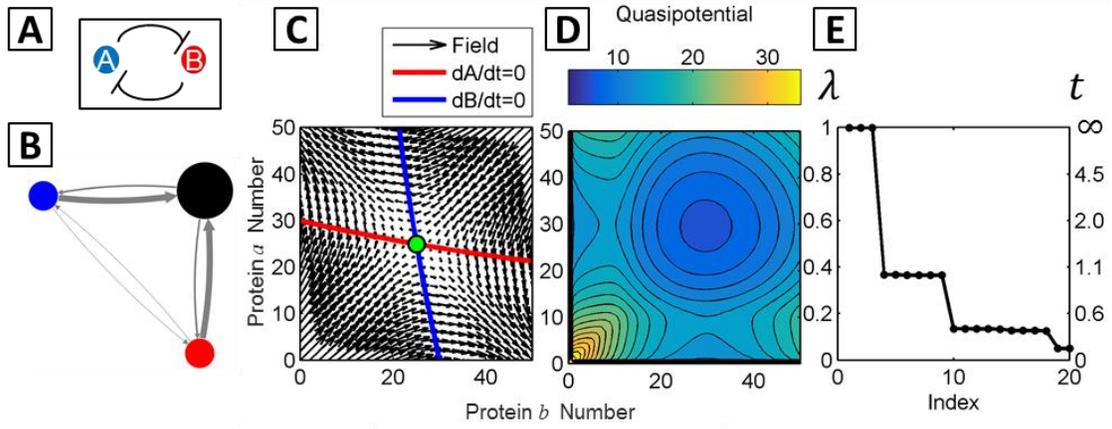

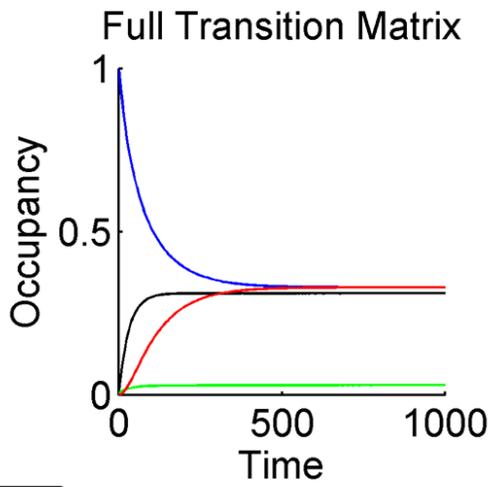
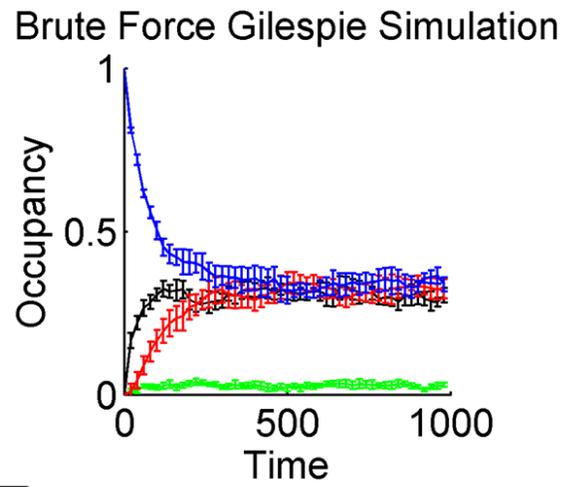
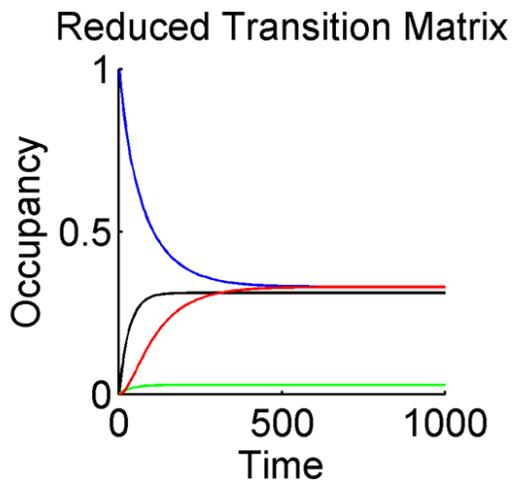
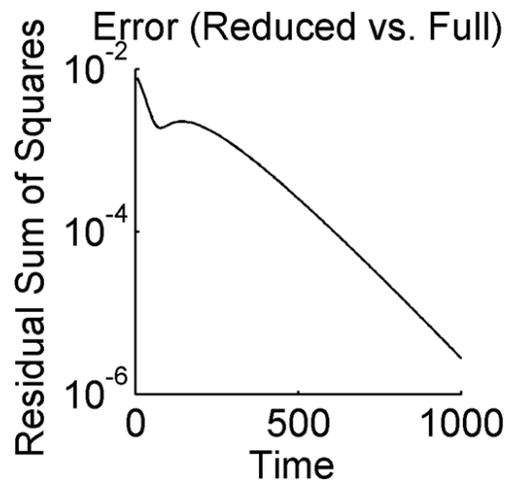

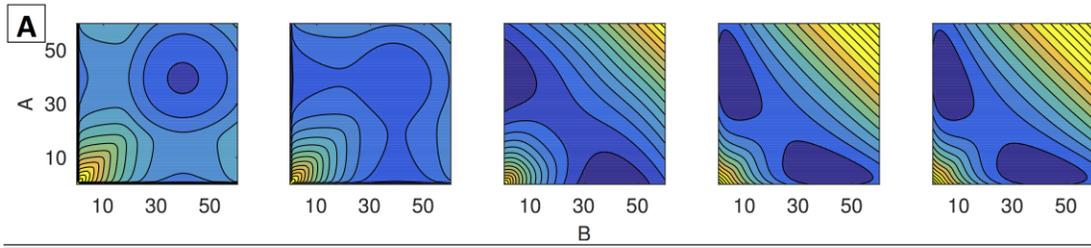
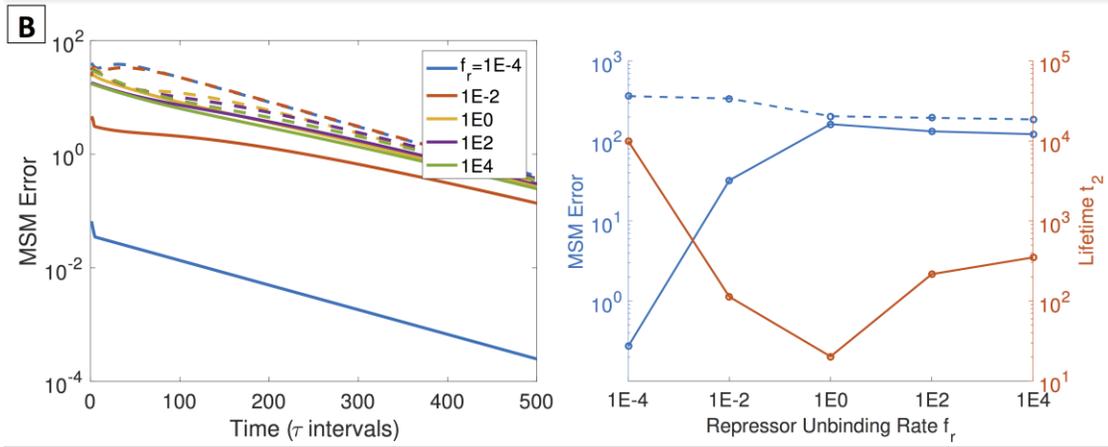
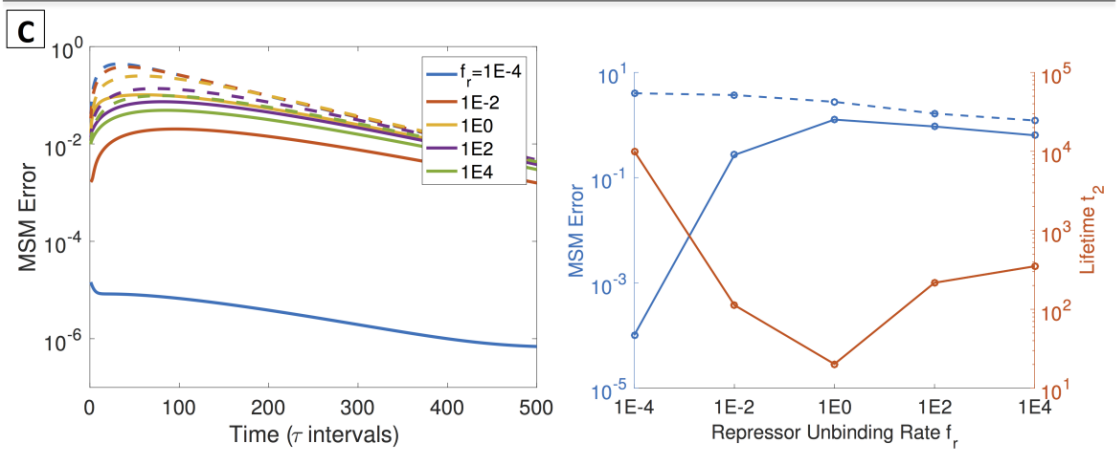

| A | Pathway Probability (%) | | | | Stationary Probability (%) | | | |
|---|---|---|---|---|---|---|---|---|
| $f_r$ | 🔵→🔴 | 🔵→🟢→🔴 | 🔵→⚫→🔴 | 🔵→⚫→🟢→🔴 | 🔵 | ⚫ | 🟢 | 🔴 |
| 5E-4 | 1.0 | 61.8 | 36.9 | 2.3E-01 | 27.4 | 1.3 | 44.0 | 27.4 |
| 1E-3 | 1.5 | 44.7 | 53.5 | 3.3E-01 | 34.5 | 3.2 | 27.8 | 34.5 |
| 5E-3 | 2.3E-01 | 13.8 | 83.4 | 5.0E-01 | 37.9 | 17.9 | 6.3 | 37.9 |
| 1E-2 | 2.4E-01 | 7.3 | 89.7 | 5.2E-01 | 33.0 | 31.1 | 2.9 | 33.0 |
| 1.5E-2 | 2.5E-01 | 4.9 | 92.1 | 5.2E-01 | 28.8 | 40.7 | 1.7 | 28.8 |

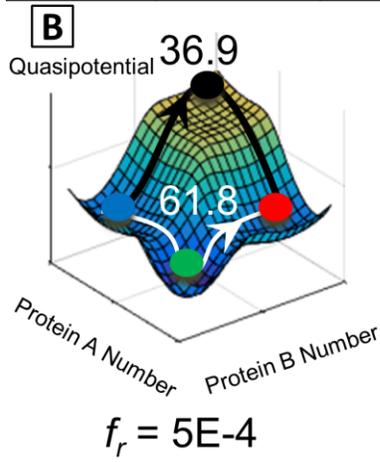
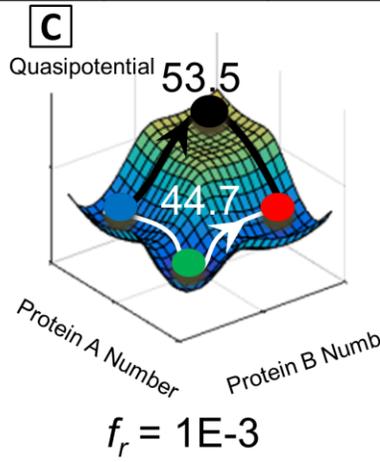
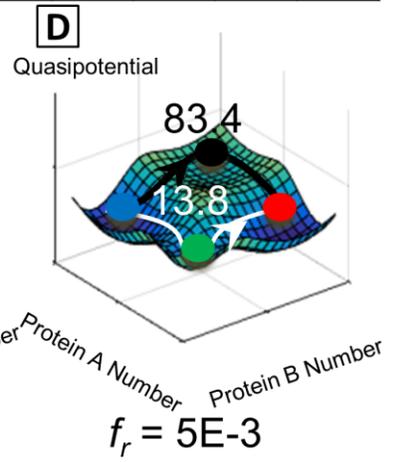

$f_r$ = 5E-4     $f_r$ = 1E-3     $f_r$ = 5E-3

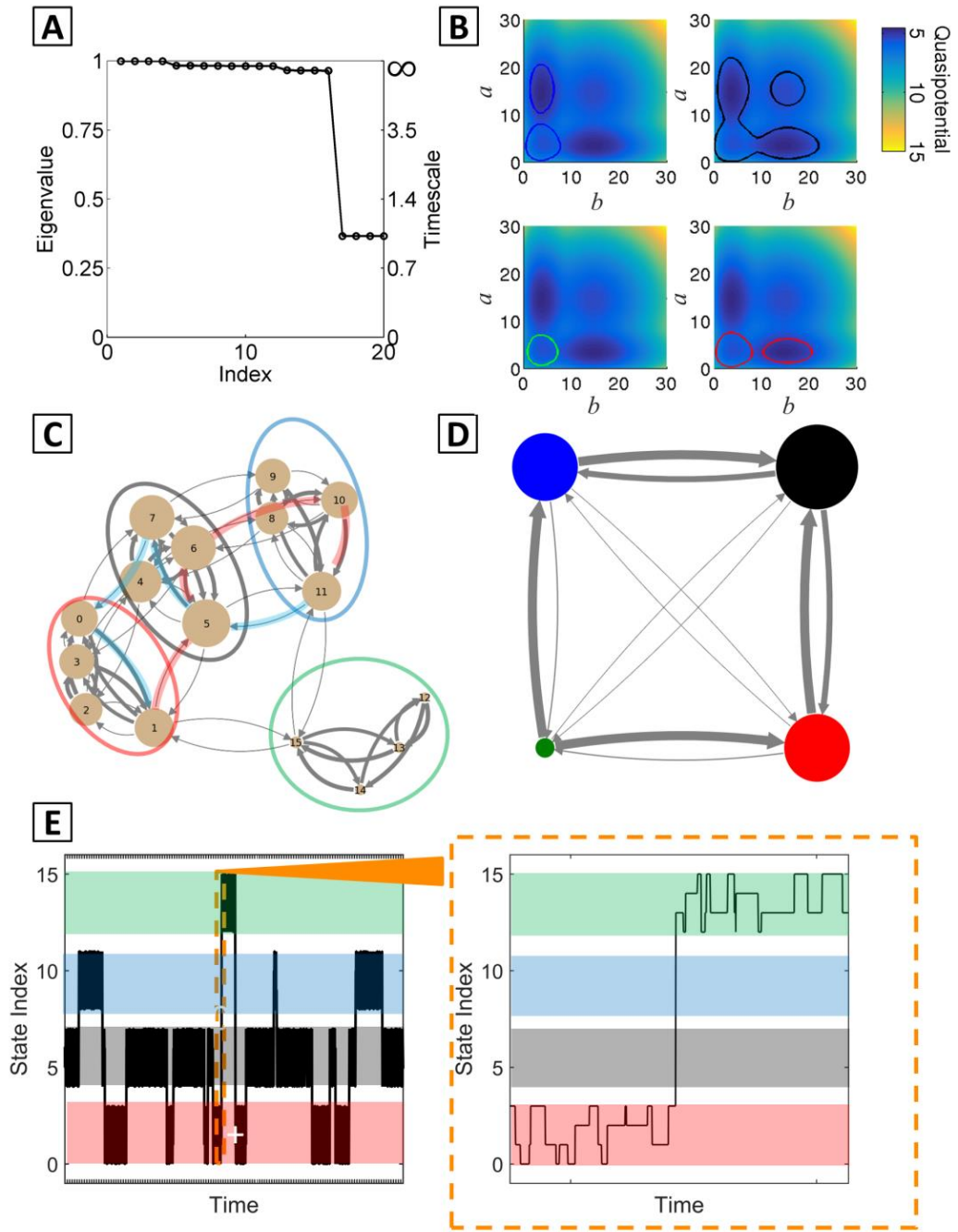